\documentclass[11pt]{article}
\usepackage{graphicx}
\usepackage{amssymb}
\def\lsim{\mathrel{\raise.3ex\hbox{$<$\kern-.75em\lower1ex\hbox{$\sim$}}}}
\def\gsim{\mathrel{\raise.3ex\hbox{$>$\kern-.75em\lower1ex\hbox{$\sim$}}}}

\begin{document}


\title{Confronting mass-varying neutrinos with MiniBooNE}

\author{V. Barger$^{1}$, D. Marfatia$^{2}$ and K. Whisnant$^{3}$\\[2ex]
\small\it $^1$Department of Physics, University of Wisconsin, Madison, WI 53706\\
\small\it $^2$Department of Physics and Astronomy, University of Kansas, Lawrence, KS 66045\\
\small\it $^3$Department of Physics and Astronomy, Iowa State University, Ames, IA 50011}

\date{}

\maketitle

\begin{abstract}

We study the proposal that mass-varying neutrinos could provide an
explanation for the LSND signal for $\bar\nu_\mu \to \bar\nu_e$
oscillations. We first point out that all positive oscillation signals
occur in matter and that three active mass-varying neutrinos are
insufficient to describe all existing neutrino data including LSND.
We then examine the possibility that a model with four mass-varying
neutrinos (three active and one sterile) can explain the LSND effect
and remain consistent with all other neutrino data. We find that such
models with a 3~+~1 mass structure in the neutrino sector may explain
the LSND data and a null MiniBooNE result for $0.10 \lsim \sin^2
2\theta_x \lsim 0.30$. Predictions of the model include a null result
at Double-CHOOZ, but positive signals for underground reactor
experiments and for $\nu_\mu \to \nu_e$ oscillations in long-baseline
experiments.

\end{abstract}

\newpage

\section{Introduction}

The LSND experiment has found evidence for $\bar\nu_\mu \to \bar\nu_e$
oscillations at the $3.3\sigma$ level~\cite{lsnd-dar, lsnd-final}, with
indications for $\nu_\mu \to \nu_e$ oscillations at lesser
significance~\cite{lsnd-final, lsnd-dif}. The combination of the LSND
data with the compelling evidence for oscillations in
solar~\cite{chlorine, superK-solar, gallium, sno},
atmospheric~\cite{kam, imb, superK-atmos, soudan2, macro},
accelerator~\cite{k2k}, and reactor~\cite{kamland} neutrino
experiments cannot be adequately explained in the standard
three-neutrino picture~\cite{not-three}.  Extensions to models with
four light neutrinos (with the extra neutrino being
sterile)~\cite{sterile} or $CPT$ violation with three
neutrinos~\cite{cptv, cptv2} have been proposed to accommodate all
neutrino data. However, in both cases, recent analyses indicate that
neither scenario provides a good description of the complete data
set~\cite{not-four, not-cptv}.\footnote{It has been
suggested~\cite{pas-song-weiler} that certain approximations in some
of these analyses have ignored small terms that may allow a better fit
to the data, but a full study has not yet been made.} The addition of
a second sterile neutrino improves the fit substantially~\cite{sorel}
in some models with more than three neutrinos, especially if there is
$CP$ violation~\cite{32CPV}.  Another possible solution is to have
$CPT$ violation in a four-neutrino model~\cite{4nuCPTV}. Neutrino
decay has also been proposed~\cite{nu-decay} as a means of explaining
the LSND data.

Mass-varying neutrinos~\cite{mavans} (MaVaNs) have been discussed as a
means for generating dark energy~\cite{dark-energy}, explaining the
LSND results~\cite{zurek} and have been shown to improve the neutrino
oscillation fit to solar neutrino data~\cite{solar-mavans}.  It is a
relevant fact that for all positive oscillation signals (solar,
atmospheric, K2K, KamLAND, and LSND) the detected neutrino travels
through matter for some or all of its path length. In all but solar
neutrino oscillations the path lengths are nearly all through Earth
matter, while in the solar case the important matter effects occur in
the sun. Therefore if MaVaNs exist, it is possible (although not
required) that the masses and mixing angles indicated by the data
could differ substantially from their vacuum values.

There are three positive oscillation signals in which the neutrino
path is primarily in the Earth's crust: KamLAND, K2K and LSND. The
latest data from K2K~\cite{k2k} now yield an allowed region for the
mass-squared difference which is roughly similar to that obtained for
the oscillations of atmospheric neutrinos and is inconsistent with the
mass scales indicated by KamLAND and LSND neutrino oscillations.  The
usual argument used to exclude three-neutrino models from solar,
atmospheric and LSND data~\cite{not-three} is that there are only two
independent mass-squared differences for three neutrinos. This
argument applies in the MaVaN scenario to the combination of KamLAND,
K2K and LSND experiments, all of which were conducted in Earth matter
with similar density (and which therefore should be subject to similar
mixings and mass scales).  Thus a three-MaVaN model is insufficient to
explain all of the data.

In this paper we explore the possibility that oscillations of three
active and one sterile MaVaN can explain all neutrino data including
LSND and MiniBooNE. Since neutrinos in the KamLAND, K2K and LSND
experiments all pass primarily through Earth matter of approximately
the same density (Earth crust), they should be consistent with the
same set of mixing angles and mass-squared differences. If a set of
oscillation parameters cannot be found that is consistent with these
three experiments, then a four-MaVaN model is not possible. If such a
set can be found, consistency with the solar, atmospheric and vacuum
neutrino data must also be realized for the model to be viable. We
study only 3~+~1 models, in which there is one mass eigenstate
well-separated from the other three, since the constraints on 2~+~2
models are much stronger~\cite{not-four,our-review}.

For simplicity we examine a minimal 3~+~1 MaVaN scenario in which
substantial MaVaN effects occur only for the $\nu_3$ and $\nu_4$
states and there is no vacuum mixing between active and sterile
neutrinos, so that MaVaN effects are solely responsible for
active-sterile mixing. We study the feasibility of such a model in
describing current data. The MiniBooNE experiment~\cite{miniboone} is
now taking data that will test the LSND oscillation parameters in the
$\nu_\mu \to \nu_e$ channel; we examine the consequences of this 3~+~1
MaVaN model for both positive and negative MiniBooNE results. We find
that viable solutions exist if MiniBooNE sees no oscillations.
Furthermore, in those solutions both the vacuum neutrino masses in the
active sector and the MaVaN mass terms need not exhibit a hierarchy,
and the mass scales for the oscillations of both atmospheric and LSND
neutrinos are generated by MaVaN effects. Solar neutrino and KamLAND
data are explained primarily by vacuum masses and mixings. Finally, we
discuss the implications of this model for future experiments.

\section{A 3~+~1 MaVaN model}

\subsection{Masses and mixings}

An element of the mass-squared matrix for MaVaNs in the vacuum eigenstate basis
can be written as
\begin{equation}
(M^2)_{ij} = (m_i\delta_{ij} - M_{ij})^2 \,,
\end{equation}
where the $m_i$ are the masses in an environment dominated by the
cosmic microwave background and the $M_{ij}$ are the density-dependent
mass terms generated by acceleron couplings to matter fields. We will
assume that the heaviest neutrinos have masses of
${\cal{O}}(0.01)$~eV in the present epoch, and that as a result of
their non-negligible velocities, the neutrino overdensity in the Milky
Way from gravitational clustering can be neglected~\cite{singh}. The
$m_i$ (which we will refer to as vacuum masses) represent the masses
of terrestrial neutrinos in laboratory experiments like those
measuring tritium beta decay~\cite{beta-decay}. We note that
cosmological bounds on the sum of neutrino masses of
${\cal{O}}(1)$~eV~\cite{bounds} are inapplicable to MaVaNs.
Consequently, the usual relationship between neutrino dark matter and
neutrinoless double beta decay~\cite{0nbb} is also rendered
inapplicable. In Ref.~\cite{zald}, it was pointed out that so long as
the acceleron does not couple to nonrelativistic neutrino eigenstates
(which is the case under consideration), neutrino dark energy is
stable. However, the stability of neutrino dark energy with the
acceleron also coupled to matter has not been studied so far.

We adopt a matter dependence of the form~\cite{solar-mavans}
\begin{equation}
M_{ij}(n_e) = M_{ij}^0 \left(n_e\over n_e^0\right)^k \,,
\label{eq:M}
\end{equation}
where $n_e$ is the electron number density in units of $N_A/$cm$^3$,
$M_{ij}^0$ are the values at some reference density $n_e^0$ and $k$
parametrizes a power law dependence of the neutrino mass on
density. In principle, $M_{ij}$ is expected to depend linearly on $n_e$ 
since the 
acceleron is assumed to evolve adiabatically and remain at the minimum of its
potential.
 We allow $k$ to deviate from
unity to emphasize that a wider range of $k$ is allowed by oscillation data.
The choice of reference density is arbitrary; we will take it
to be that of the Earth's
crust, $n_e^0 \simeq 1.5$. Implicit in the form of Eq.~(\ref{eq:M}) is
the assumption that the neutrino number density has a negligible
effect on neutrino masses. Thus, it applies only in the current epoch
when the cosmic neutrino background number density
(${\cal{O}}(10^{-12})$ eV$^3$) is tiny. At earlier epochs, the
neutrino number density is orders of magnitude larger and must be
taken into account. For example, in the era of Big Bang
Nucleosynthesis, the neutrino number density is ${\cal{O}}(10^{30})$
eV$^3$. Moreover, we have no reason to expect the $M_{ij}^0$ to be
unaltered at earlier epochs since the acceleron-matter couplings may
vary with redshift. 

For simplicity we will study a MaVaN model in which MaVaN effects
occur only for $\nu_3$ and $\nu_4$ and there is no vacuum
mixing between active and sterile neutrinos. The rationale for this
choice is that if MaVaN effects are to be responsible for generating
the LSND mass-squared difference and for mixing between active and
sterile neutrinos in the Earth, then if they also involve the two
lightest neutrinos it would be difficult to obtain the proper
mass-squared difference for solar neutrinos. Then the evolution
equations in the flavor basis for MaVaN oscillations in matter at the
reference density is given by
\begin{equation}
i {d\over dL} \pmatrix{\nu_e \cr \nu_\mu \cr \nu_\tau \cr \nu_s} =
{1\over2E_\nu} \left[ U M^2 U^\dag
+ A\pmatrix{1 & 0 & 0 & 0 \cr 0 & 0 & 0 & 0\cr 0 & 0 & 0 & 0\cr 0 & 0 & 0 & r}
\right]
\pmatrix{\nu_e \cr \nu_\mu \cr \nu_\tau \cr \nu_s} \,,
\label{eq:prop}
\end{equation}
where 
\begin{equation}
M^2 = \pmatrix{m_1^2 & 0 & 0 & 0 \cr
0 & m_2^2 & 0 & 0 \cr 0 & 0 & (m_3 - M_{33})^2 & M_{34}^2 \cr
0 & 0 & M_{34}^2 & (m_4 - M_{44})^2} \,,
\label{eq:M2a}
\end{equation}
$U$ is the neutrino mixing matrix that connects the flavor eigenstates
with the mass eigenstates in vacuum, given by
\begin{equation}
U = \pmatrix{U_{e1} & U_{e2} & U_{e3} & 0 \cr
U_{\mu1} & U_{\mu2} & U_{\mu3} & 0 \cr
U_{\tau1} & U_{\tau2} & U_{\tau3} & 0 \cr
0 & 0 & 0 & 1}
= \pmatrix{c_x c_s & c_x s_s & s_x & 0 \cr
-c_a s_s - s_a c_s s_x
& c_a c_s - s_a s_s s_x
& c_x s_a & 0 \cr
s_a s_s - c_a c_s s_x
& - s_a c_s - c_a s_s s_x
& c_x c_a & 0 \cr
0 & 0 & 0 & 1} \,,
\label{eq:U}
\end{equation}
and $s_j$ and $c_j$ denote $\sin\theta_j$ and $\cos\theta_j$,
respectively, for $j= s, a, x$. The angle $\theta_s$ represents the
usual mixing for solar neutrinos, $\theta_a$ the mixing for
atmospheric neutrinos, and $\theta_x$ the mixing of $\nu_e$ with
$\nu_\mu$ in atmospheric and long-baseline experiments. The charged-current
amplitude for $\nu_e-e$ forward scattering in matter is~\cite{A}
\begin{equation}
A = 2\sqrt2 G_F n_e E_\nu = 1.52\times10^{-7}{\rm eV}^2 \, n_e 
E_\nu({\rm MeV})\,.
\end{equation}
There is also a $\nu-e$ neutral-current forward scattering contribution
for all active neutrinos given by $-Ar$, where $r = n_n/(2n_e)$ and
$n_n$ is the neutron number density in units of $N_A/$cm$^3$, which is
equivalent to a $+Ar$ term for the sterile neutrino (see Eq.~\ref{eq:prop}).

It is convenient to parametrize the mass-squared matrix in matter in
terms of its eigenvalues $M_i^2$ and the mixing angle $\theta$ that
diagonalizes it,
\begin{equation}
M^2 =
\pmatrix{1 & 0 & 0 & 0 \cr 0 & 1 & 0 & 0 \cr
0 & 0 & \cos\theta & -\sin\theta \cr 0 & 0 & \sin\theta & \cos\theta}
\pmatrix{M_1^2 & 0 & 0 & 0 \cr 0 & M_2^2 & 0 & 0 \cr
0 & 0 & M_3^2 & 0 \cr 0 & 0 & 0 & M_4^2}
\pmatrix{1 & 0 & 0 & 0 \cr 0 & 1 & 0 & 0 \cr
0 & 0 & \cos\theta & \sin\theta \cr 0 & 0 & -\sin\theta & \cos\theta} \,;
\label{eq:M2b}
\end{equation}
then the mass-squared differences are $\delta M_{ij}^2 = M_i^2 -
M_j^2$. If the ordinary matter term can be ignored (which will be
approximately true for all but solar neutrinos), the neutrino mixing
between the mass eigenstates in matter and the flavor eigenstates is
\begin{equation}
V = U \pmatrix{1 & 0 & 0 & 0 \cr 0 & 1 & 0 & 0 \cr
0 & 0 & \cos\theta & -\sin\theta \cr 0 & 0 & \sin\theta & \cos\theta}
= \pmatrix{U_{e1} & U_{e2} & c U_{e3} & -s U_{e3} \cr
U_{\mu1} & U_{\mu2} & c U_{\mu3} & -s U_{\mu3} \cr
U_{\tau1} & U_{\tau2} & c U_{\tau3} & -s U_{\tau3} \cr
0 & 0 & s & c} \,,
\label{eq:V}
\end{equation}
where $s \equiv \sin\theta$ and $c \equiv \cos\theta$. For solar
neutrinos, where the ordinary matter terms are important, $V$ cannot
easily be written in closed form. Since the MaVaN terms do not affect
the first and second generations, $M_1 = m_1$ and $M_2 = m_2$.

In 3~+~1 models there is one neutrino mass well-separated from the
others by the LSND mass-squared difference ($\delta m^2_L$), and the
sterile neutrino couples strongly only to the isolated state. The
atmospheric and solar mass-squared differences will be denoted by
$\delta m^2_a$ and $\delta m^2_s$, respectively. There are four
possible mass spectra in 3~+~1 models, depending on whether the
isolated state is above or below the others, and whether the other
three neutrino states have a normal ($\delta m^2_a >0$) or inverted
($\delta m^2_a <0$) mass hierarchy. We only consider the case with
$M_4 > M_1, M_2, M_3$ and normal hierarchy, which implies $\delta
M^2_{43} \simeq \delta M^2_{42} \simeq \delta M^2_{41} = \delta m^2_L
\gg \delta M^2_{32} \simeq \delta M^2_{31} = \delta m^2_a \gg \delta
M^2_{21} = \delta m^2_s$.

For now we take $m_1 = M_1 = 0$, so that $m_2^2 = M_2^2 = \delta
m^2_s$, $M_3^2 = \delta m^2_a$ and $M_4^2 = \delta m^2_L$ (we discuss
the case $m_1 \ne 0$ in Sec.~4). We will also set $m_4 = 0$, {\it i.e.}, the
sterile neutrino is massless in vacuum. In our illustration we take
$\sin^22\theta_s = 0.80$ and $\sin^22\theta_a = 1.00$, and allow
$\theta_x$ to vary. For the remaining parameters ($m_3, M_{33},
M_{34}, M_{44}$) we have the following relations that follow from the
diagonalization of $M^2$ (Eqs.~\ref{eq:M2a} and \ref{eq:M2b}):
\begin{eqnarray}
(m_3 - M_{33})^2 = c^2 M_3^2 + s^2 M_4^2 \,,
\label{eq:m33}\\
M_{44}^2 = s^2 M_3^2 + c^2 M_4^2 \,,
\label{eq:m44}\\
M_{34}^2 = s c (M_3^2 - M_4^2) \,.
\label{eq:m34}
\end{eqnarray}
We will also assume that the vacuum masses $m_2$ and $m_3$ are much smaller
than the MaVaN parameters in the Earth's crust ($M_{33}$, $M_{44}$ and
$M_{34}$), which will be justified by our numerical results in Sec.~3.

In order to determine the allowed MaVaN parameters, we will use the
positive oscillation results from KamLAND, K2K and LSND. Since these
experiments were conducted in the Earth's crust, the mixing matrix $V$
for all of them should be nearly the same; we will consider it as the
same matrix for all three experiments and attempt to determine the
parameters from the combined data. Since the baselines for these
experiments are all 250~km or less, ordinary matter effects due to
coherent forward scattering are small and the $A$ term in
Eq.~(\ref{eq:prop}) can be ignored to a good approximation. Therefore
the mass-squared differences relevant for oscillations in these
experiments are $\delta M^2_{ij}$ and the mixing matrix is $V$ in
Eq.~(\ref{eq:V}).

There are also constraints from other experiments, but only those
experiments which were conducted primarily in Earth matter are
relevant for the matrix that describes the results of KamLAND, K2K and
LSND. The CHOOZ~\cite{chooz} reactor constraint on $\bar\nu_e \to
\bar\nu_e$ oscillations at the $\delta m^2_a$ scale ($L/E_\nu \simeq
250$~m/MeV) does not apply since the neutrino path in the CHOOZ
experiment was primarily in air; it is instead replaced by the weaker
Palo Verde~\cite{palo-verde} constraint at a somewhat smaller
$L/E_\nu$ value.  Similarly, bounds from the Bugey reactor
experiment~\cite{bugey} at the $\delta m^2_L$ scale ($L/E_\nu \simeq
25$~m/MeV) must be replaced by the considerably weaker bounds from
Gosgen~\cite{gosgen} and Krasnoyarsk~\cite{krasnoyarsk}. The
CDHSW~\cite{cdhsw} bound on $\nu_\mu \to \nu_\mu$ oscillations at the
$\delta m^2_L$ scale applies to the Earth matter case, as the
neutrino path was approximately 90\% in matter in this
experiment~\cite{cdhsw-mat}. Once the MaVaN parameters have been
determined from the Earth crust data, consistency with the
atmospheric, solar and vacuum data can be checked.

\subsection{Oscillation formulas}

In this section we list the oscillation probabilities in the limit
that the ordinary matter effect can be ignored, except for solar
neutrino oscillations. The relevant oscillation probabilities in the
leading oscillation are approximately
\begin{eqnarray}
P(\bar\nu_\mu \to \bar\nu_e)_{\rm LSND} &\simeq&
4 s^4 |U_{\mu3}|^2 |U_{e3}|^2 \sin^2\Delta_L \,,
\label{eq:PmeL}\\
P(\nu_\mu \to \nu_\mu)_{\rm CDHSW} &\simeq&
1 - 4 s^2 |U_{\mu3}|^2 (1 - s^2 |U_{\mu3}|^2) \sin^2\Delta_L \,,
\label{eq:PmmL}\\
\bar P(\bar\nu_e \to \bar\nu_e)_{\rm Gosgen} &\simeq&
1 - 4 s^2 |U_{e3}|^2 (1 - s^2 |U_{e3}|^2) \sin^2 \Delta_L \,,
\label{eq:PeeL}
\end{eqnarray}
where $\Delta_L$ is the largest of the usual oscillation arguments
$\Delta_j = \delta m^2_j L/(4E_\nu)$ for $j = L$, $s$, $a$.

The relevant oscillation probabilities at the first subleading
scale ($\delta m^2_a$) are approximately
\begin{eqnarray}
P(\nu_\mu \to \nu_\mu)_{\rm K2K} &\simeq&
1 - 2 s^2 |U_{\mu3}|^2 (1 - s^2 |U_{\mu3}|^2)
- 4 c^2 |U_{\mu3}|^2 (1 - |U_{\mu3}|^2) \sin^2\Delta_a
\label{eq:Pmma}\\
P(\nu_\mu \to \nu_e)_{\rm K2K} &\simeq&
2 s^4 |U_{e3}|^2 |U_{\mu3}|^2 +
4 c^2 |U_{e3}|^2 |U_{\mu3}|^2 \sin^2\Delta_a \,,
\label{eq:Pmea}\\
P(\nu_\mu \to \nu_s)_{\rm K2K} &\simeq&
2 s^2 c^2 |U_{\mu3}|^2 \,,
\label{eq:Pmsa}\\
P(\bar\nu_e \to \bar\nu_e)_{\rm Palo Verde} &\simeq&
\sin^22\theta_x \left[ {1\over2} s^4  + c^2 \sin^2\Delta_a \right] \,,
\label{eq:Peea}
\end{eqnarray}
where we have averaged over the leading oscillation scale. We do not
consider $CP$ violation, so the oscillation probabilities for
neutrinos and anti-neutrinos are the same.

At the smallest scale ($\delta m^2_s$) we have
\begin{eqnarray}
P(\bar\nu_e \to \bar\nu_e)_{\rm KamLAND} &\simeq&
1 - 2 |U_{e3}|^2 (1 - |U_{e3}|^2) - 2 s^2 c^2 |U_{e3}|^4
- 4 |U_{e1}|^2 |U_{e2}|^2 \sin^2\Delta_s \,,
\label{eq:Pees}
\end{eqnarray}
where we have averaged over the oscillations of the two higher $\delta
m^2$ scales. For solar neutrinos, for the large mixing angle (LMA)
solution with adiabatic propagation we have
\begin{equation}
P(\nu_e \to \nu_e) = \sum_{j=1}^4 |U_{ej}|^2|V_{ej}^0|^2 \,,
\label{eq:solar}
\end{equation}
where $V^0$ is the corresponding value of $V$
at the point in the sun where the neutrino is created
($n_e \simeq 80$). The fraction of solar neutrinos that oscillate to
sterile neutrinos is given by
\begin{equation}
P(\nu_e \to \nu_s) = \sum_{j=1}^4 |U_{ej}|^2 |V_{sj}^0|^2 \,.
\label{eq:solar-sterile}
\end{equation}

\section{Model constraints}

\subsection{Fits to Earth crust data}

As discussed above, KamLAND, K2K and LSND should all be described with
the $\delta M^2_{ij}$ and mixing $V$ that exist in the Earth's crust
($n_e \simeq 1.5$). KamLAND data~\cite{kamland} give $\delta m^2_s
\simeq 8\times10^{-5}$~eV$^2$ with oscillation amplitude 0.8 and the
K2K results~\cite{k2k} imply $\delta m^2_a \simeq 2.8\times
10^{-3}$~eV$^2$ with maximal mixing (both in two-neutrino fits).  In a
three-neutrino model, if the mixing angle $\theta_x$ vanishes, then
these two oscillations decouple from one another~\cite{bw-global} and
the two-neutrino fits may be used directly. However, the amplitude of
the LSND oscillation is (from Eq.~\ref{eq:PmeL})
\begin{equation}
\sin^22\theta_L = s^4 s_a^2 \sin^22\theta_x \,,
\label{eq:sin22thL}
\end{equation}
so a non-zero $\theta_x$ is required to generate an LSND signal.  We
will examine solutions with small sterile mixing ($\sin^2\theta \ll
1$), so that reactor constraints at the $\delta m^2_a$ scale are
similar to the usual three-neutrino case (see
Eq.~\ref{eq:Peea}). Similarly, the oscillation probabilities for
KamLAND and solar neutrinos are approximately the same as in the
three-neutrino case (see Eq.~\ref{eq:Pees} and Section~3.3). There are
upper bounds on $\theta_x$ from three-neutrino
fits~\cite{fogli}: $\sin^22\theta_x \le 0.19~(0.28)$ from
solar and KamLAND data and $\sin^22\theta_x \le 0.16~(0.23)$ from
CHOOZ, atmospheric and K2K data, where the bounds are at the
$2\sigma$ ($3\sigma$) level. However, for MaVaNs traveling through the
Earth's crust the CHOOZ bound is no longer applicable and must be
replaced by the Palo Verde constraint, which is less restrictive by
approximately a factor of two at the $\delta m^2_a$ indicated by K2K
and atmospheric neutrinos ($\approx 2-3 \times 10^{-3}$~eV$^2$).
Therefore, a value of $\sin^22\theta_x$ as large as $\simeq 0.3$ would
appear to be allowed at the $3\sigma$ level with the CHOOZ constraint
removed.

Finally, it is well-known that a combination of reactor and
accelerator constraints disfavor the standard 3~+~1
model~\cite{not-four,4nu-constraints}. In our 3~+~1 MaVaN model, the
strong bounds from the Bugey reactor are replaced by the considerably
weaker bounds (by approximately a factor of three) from Gosgen and
Krasnoyarsk. In the region of interest ($\delta m^2_L \simeq
1$~eV$^2$) the upper bound on the $\nu_\mu \to \nu_\mu$ oscillation
amplitude from CDHSW is about 0.1; the corresponding bound on the
$\bar\nu_e \to \bar\nu_e$ oscillation amplitude from
Gosgen/Krasnoyarsk is also about 0.1. From Eqs.~(\ref{eq:PmmL}) and
(\ref{eq:PeeL}) we see that in this model the oscillation amplitude
for Gosgen/Krasnoyarsk is smaller than that for CDHSW by a factor
$|U_{e3}/U_{\mu3}|^2 = \tan^2\theta_x/s_a^2$ (for small $\theta$),
which for $\sin^22\theta_x \le 0.3$ and $\theta_a = \pi/4$ is of order
0.15 or less. Therefore if the CDHSW bound is satisfied, then the
Gosgen/Krasnoyarsk constraints are automatically satisified, and we
need to consider only the effect of the CDHSW bound on the model
parameters.

\begin{figure}[th]
\centering\leavevmode
\includegraphics[width=4.5in]{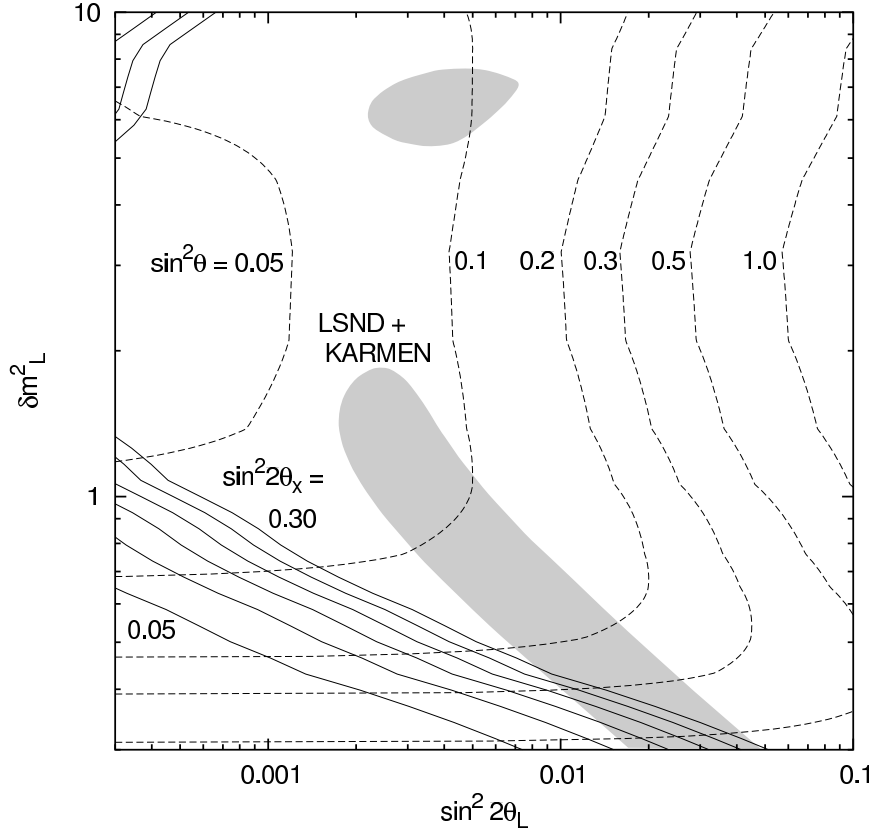}
\caption[]{Regions in $\delta m^2_L$-$\sin^22\theta_L$ space
accessible to the 3~+~1 MaVaN model discussed in this paper. The solid
(dashed) curves show the upper bound on the LSND amplitude 
$\sin^22\theta_L$ from
CDHSW for given values of $\sin^22\theta_x$ 
($\sin^2 \theta$, the mixing between the third and fourth mass 
eigenstates due to MaVaN effects).  
The intersections of the dashed curves with the solid
curves show the values of $\sin^2\theta$ and $\sin^22\theta_x$ 
required to achieve the maximum LSND amplitudes allowed by CDHSW.
The shaded regions show the current 99\%~C.L. region allowed by a joint
analysis~\cite{LSND-KARMEN} of LSND and KARMEN data.
\label{fig:fig1}}
\end{figure}

In Fig.~\ref{fig:fig1} we show regions in LSND amplitude
$\sin^22\theta_{L}$ and mass-squared difference $\delta m^2_L$
accessible to the model and the constraints from data. The solid
curves are the upper bounds on $\sin^22\theta_L$ allowed by the 90\%~C.L.
CDHSW constraint for several values of $\theta_x$. The intersections of
the dashed curves with the solid curves show
the values of $\sin^2\theta$ required to achieve those upper bounds;
{\it i.e.}, $s^4 = 2\sin^22\theta_L/\sin^22\theta_x$ from
Eq.~(\ref{eq:sin22thL}). The shaded regions show the current
99\%~C.L. allowed region from a joint analysis~\cite{LSND-KARMEN} of LSND
and KARMEN~\cite{karmen} data. For $\sin^22\theta_x \le 0.30$,
$\sin^2\theta \gsim 0.23$ is required to be consistent with the
LSND/KARMEN 99\%~C.L. allowed region. However, as we will show in the next
section, such large values of $\theta$ are not compatible with
atmospheric neutrino data.

\subsection{Atmospheric neutrinos}

Expressions for the oscillation probabilities for atmospheric
neutrinos are similar to those for K2K
(Eqs.~\ref{eq:Pmma}-\ref{eq:Pmsa}), except that the values for $\delta
M^2$ and $V$ vary as the electron number density along the neutrino
path varies (there is also an additional matter effect for $\nu_\mu
\to \nu_e, \nu_s$ oscillations for the higher-energy atmospheric
neutrinos, which we ignore). We will call the varying mass and mixing
parameters for atmospheric neutrinos $\tilde M_3^2$, $\tilde M_4^2$
and $\tilde \theta$. These quantities obey relations similar to
Eqs.~(\ref{eq:m33})-(\ref{eq:m34})
\begin{eqnarray}
(m_3 - M_{33} r^k)^2 = \tilde c^2 \tilde M_3^2 + \tilde s^2 \tilde M_4^2 \,,
\label{eq:m33t}\\
M_{44}^2 r^{2k} = \tilde s^2 \tilde M_3^2 + \tilde c^2 \tilde M_4^2
\,,
\label{eq:m44t}\\
M_{34}^2 r^{2k} = \tilde s \tilde c (\tilde M_3^2 - \tilde M_4^2) \,.
\label{eq:m34t}
\end{eqnarray}
with $M_3$, $M_4$ and $\theta$ replaced by their tilde counterparts
and with the MaVaN parameters $M_{33}$, $M_{34}$ and $M_{44}$
multiplied by the factor $r^k$, where $r$ is the ratio of the average
matter density for a given path compared to the density of the Earth's
crust; {\it e.g.}, for a path through the center of the Earth, $r \sim 3$.
For small $m_3$ (the solutions we are investigating) the value of the
varying sterile mixing in the Earth, $\tilde\theta$, is very similar
to the sterile mixing in the crust, $\theta$; this can be seen by
comparing the expressions for $\theta$ and $\tilde\theta$
\begin{equation}
\tan 2\theta = {2 M_{34}^2\over (m_3 - M_{33})^2 - M_{44}^2}
\qquad , \qquad
\tan 2\tilde\theta = {2 M_{34}^2 r^{2k} \over
(m_3 - M_{33} r^k)^2 - M_{44}^2 r^{2k}} \,.
\end{equation}
In the limit that $|m_3| \ll |M_{33}|$, it is evident that $\tilde\theta
\simeq \theta$. Therefore if $\theta$ is small then $\tilde \theta$
will also be small, and sterile mixing will not upset the atmospheric
neutrino fits. In this same limit the size of $\tilde M^2_4$, {\it i.e.},
the largest varying oscillation mass scale in the Earth, is
approximately $r^{2k} \delta m^2_L$.

For $\theta_a = \pi/4$, the amplitude for oscillation to sterile
neutrinos is given approximately by $\tilde s^2 c_x^2$ (from the tilde
equivalent to Eq.~\ref{eq:Pmsa}); these oscillations occur at the
leading mass scale ($\gsim 1$~eV$^2$) and oscillations with amplitude
of order $\tilde s^2 c_x^2$ are seen for downward as well as upward
neutrinos. Since no large
suppression of downward events is observed~\cite{superK-atmos}, a
value of $\sin^2\tilde\theta$ (and hence also $\sin^2\theta$) greater
than about 0.10 is disfavored. As noted in the previous section, if
$\sin^22\theta_x \le 0.30$, then the sterile mixing angle must satisfy
$\sin^2\theta \gsim 0.23$ to obtain a value for the LSND amplitude
consistent with the 99\%~C.L. allowed region from LSND and
KARMEN. Therefore, if MiniBooNE (in which the neutrino path is
primarily in Earth crust) were to confirm the LSND/KARMEN
99\%~C.L. allowed region, our 3~+~1 MaVaN model would be
disfavored.\footnote{We note that this does not necessarily rule out a
more general four-neutrino MaVaN model.}

\begin{figure}[th]
\centering\leavevmode
\includegraphics[width=4.5in]{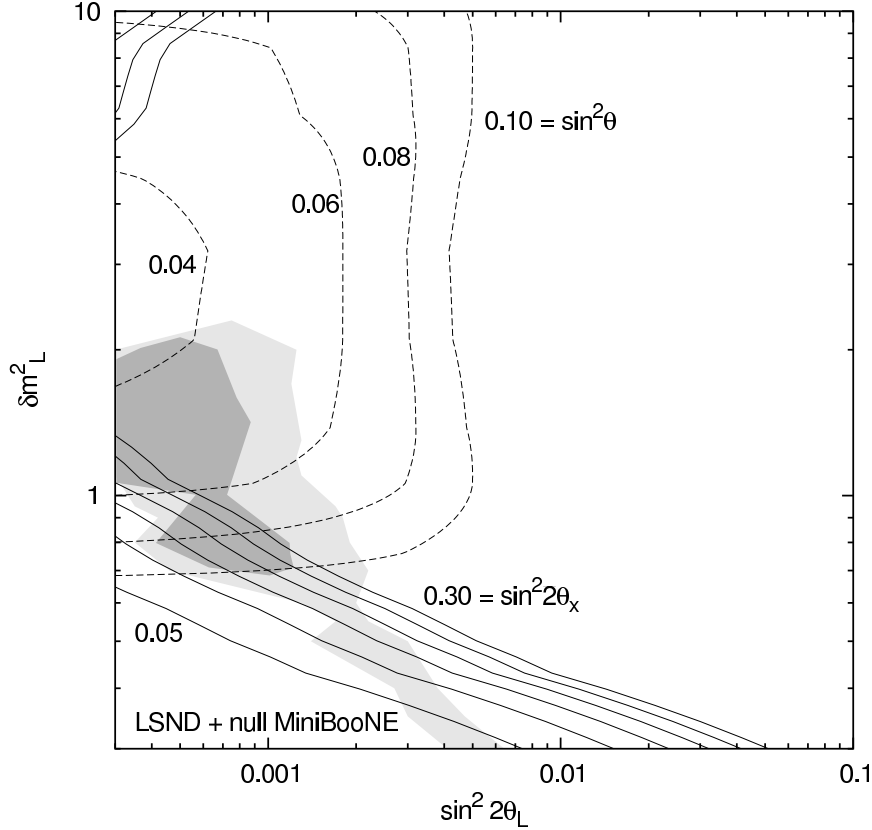}
\caption[]{Same as Fig.~\ref{fig:fig1}, except that the lighter (darker)
shaded region shows the region that would be allowed at 99.5\%~CL if
MiniBooNE sees no signal after running with $5\times10^{20}$
($10^{21}$)~P.O.T.
\label{fig:fig2}}
\end{figure}

The situation changes if MiniBooNE reports a null result. The lighter
shaded region in Fig.~\ref{fig:fig2} shows the region that would be
allowed (at 99.5\%~C.L.) by a combination of LSND and a null MiniBooNE
result with $5\times10^{20}$ protons on target
(P.O.T.)~\cite{null-miniboone}. This LSND/null-MiniBooNE region is
shifted to smaller $\sin^22\theta_L$ compared to the LSND/KARMEN
region in Fig.~\ref{fig:fig1}. This shift occurs because a null
MiniBooNE result would be in conflict with LSND, and a combined fit to
the two experiments essentially results in a weighted average of the
two oscillation probabilities (0.25\% from LSND and 0\% from
MiniBooNE, respectively). Since LSND and MiniBooNE would be in
conflict, no region is allowed below the
98\%~C.L.~\cite{null-miniboone}.  As is evident from
Fig.~\ref{fig:fig2}, this new allowed region could comfortably be
explained by our 3~+~1 MaVaN model with both $\sin^22\theta_x \lsim
0.30$ and $\sin^2\theta \lsim 0.10$, in the region near $\delta m^2_L
\sim 1$~eV$^2$ and $\sin^22\theta_L \simeq 0.0003 - 0.001$.  An
increase to $10^{21}$~P.O.T. in MiniBooNE decreases the size of the
combined allowed region but does not eliminate it (see the darker
shaded region in Fig.~\ref{fig:fig2}); there is no region below the
98.8\%~C.L. with $10^{21}$~P.O.T.~\cite{null-miniboone}. The smallest
value of $\sin^22\theta_x$ consistent with the LSND/null-MiniBooNE
region is about 0.10 if $\sin^2\theta$ is not allowed to be larger
than about 0.10, and $\sin^2\theta$ must be larger than about 0.04;
see Fig.~\ref{fig:fig2}.

So far we have considered only the effects of sterile mixing on
atmospheric neutrino oscillations, which are more or less independent
of the exact neutrino masses in matter, since the effects primarily
depend on the size of $\tilde\theta$ and not the actual mass-squared
differences. However, the mass-varying mass-squared difference that
drives the oscillations of atmospheric neutrinos ($\tilde M_3$) will
be different for different neutrino paths through the Earth. We need
to show that the model can simultaneously give the correct $\delta
m^2$ for K2K in the crust and for atmospheric neutrinos.

Although a detailed analysis of atmospheric neutrinos would be
required to determine the precise effects of the density profiles on
the allowed regions of the parameters, some semi-quantitative
statements can be made. Since the largest oscillation signal occurs
for upward events, which pass through the core, as a first
approximation we consider a path through the center of the Earth as
representative of the atmospheric data; the average matter density for
these events is $r \simeq 3$ times the density in the Earth's crust.
There are then a total of six potential observables from
Eqs.~(\ref{eq:m33})-(\ref{eq:m34}) and (\ref{eq:m33t})-(\ref{eq:m34t}):
$M_3^2, M_4^2$ and $\theta$ (from K2K and LSND), and $\tilde M_3^2,
\tilde M_4^2$ and $\tilde\theta$ (from atmospheric neutrinos).

Since the sterile mixing angle $\tilde\theta$ for atmospheric
neutrinos which pass through the core is very similar to $\theta$ in
the crust, its exact value is unimportant, as long as it is
small. Furthermore, the value of the largest mass scale for
atmospheric neutrinos which pass through the core, $\tilde M_4^2$, is
not determined by data since no such oscillations are observed.
Therefore the only relevant observable from the upward atmospheric
data is the effective mass-squared difference $\tilde M_3^2$; we can
rewrite Eqs.~(\ref{eq:m33t})-(\ref{eq:m34t}) in terms of $\tilde M_3^2$
as
\begin{equation}
M_{34}^4 r^{4k} + \tilde M_3^2 [(m_3 - M_{33}r^k)^2 +
M_{44}^2 r^{2k}] = (m_3 - M_{33} r^k)^2 M_{44}^2 r^{2k} + \tilde M_3^4 \,.
\label{eq:tilde}
\end{equation}
There are four free parameters in Eq.~(\ref{eq:M2a}): $m_3$, $M_{33}$,
$M_{34}$ and $M_{44}$. Equations~(\ref{eq:m33})-(\ref{eq:m34}) and
(\ref{eq:tilde}) can be used to determine these four parameters using
the oscillation data as follows: $M_3^2 = \tilde M_3^2 = \delta m^2_a$
(assuming K2K and upward atmospheric neutrinos have the same $\delta
m^2$) and $M_4^2 = \delta m^2_L$, where particular values for the
vacuum mixing $U$ and sterile mixing angle in the crust $\theta$ are
also used as inputs. Once $m_3, M_{33}, M_{44}$ and $M_{34}$ are
determined, the size of the LSND amplitude in Eq.~(\ref{eq:sin22thL}) can be
checked for consistency with the LSND result.

Using $\delta m^2_a = 2.0\times10^{-3}$~eV$^2$ for both K2K and
atmospheric neutrinos, Table~\ref{tab:one} shows the maximum value of
$\sin^2\theta$ allowed by CDHSW, the corresponding LSND amplitude
$\sin^22\theta_L$, and the MaVaN parameters $m_3, M_{33}, M_{34}$ and
$M_{44}$ for given values of $\delta m^2_L$ and $\sin^22\theta_x$
with $k=1$ and $r=3$. We see that the $\sin^22\theta_L$ values are in
the range allowed by the combined LSND/null-MiniBooNE fit (see
Fig.~\ref{fig:fig2}). The value of $m_3$ is small, about 0.006~eV,
which confirms our assumption that $|m_3| \ll |M_{33}|$. Since the value of
$m_2$ is $\sqrt{\delta m^2_s} \simeq 0.009$~eV, then $m_3 \sim m_2$,
{\it i.e.}, all of the non-zero vacuum masses have similar size. Likewise,
all the MaVaN parameters $M_{33}, M_{34}$ and $M_{44}$ have similar
size, of ${\cal O}(1)$~eV. Therefore no hierarchies in either the vacuum
masses of the active neutrinos 
or MaVaN parameters are required to achieve the appropriate
masses and mixings.

\begin{table}
\caption{Maximum value of $\sin^2\theta$ allowed by CDHSW, the
corresponding LSND amplitude $\sin^22\theta_L$, and the mass
parameters $m_3, M_{33}, M_{34}$ and $M_{44}$, for several sets of
values for $\delta m^2_L$ and $\sin^22\theta_x$.  The density
exponent factor used in Eq.~(\ref{eq:M}) is $k=1$, and the ratio of
core path density to crust path density used in Eq.~(\ref{eq:tilde}) is
$r=3$.
\label{tab:one}}
\begin{center}\begin{tabular}{rrr|r|rrrr}
\hline
\hline
$\delta m^2_L$ & $\sin^22\theta_x$ & $(\sin^2\theta)_{max}$
& $\sin^22\theta_L$ & $m_3$ & $M_{33}$ & $M_{34}$ & $M_{44}$ \\
(eV$^2$) & & & & (eV) & (eV) & (eV) & (eV) \\
\hline
\hline
1.0 & 0.30 & 0.063 & 5.9$\times10^{-4}$ & 0.0056 & $-$0.249 & 0.492 & 0.968 \\ 
1.0 & 0.20 & 0.061 & 3.7$\times10^{-4}$ & 0.0057 & $-$0.245 & 0.489 & 0.969 \\
\hline
0.9 & 0.30 & 0.074 & 8.2$\times10^{-4}$ & 0.0055 & $-$0.256 & 0.485 & 0.913 \\ 
0.9 & 0.15 & 0.071 & 3.8$\times10^{-4}$ & 0.0056 & $-$0.251 & 0.480 & 0.914 \\
\hline
0.8 & 0.30 & 0.085 & 1.1$\times10^{-3}$ & 0.0056 & $-$0.259 & 0.472 & 0.856 \\ 
0.8 & 0.15 & 0.081 & 4.9$\times10^{-4}$ & 0.0057 & $-$0.252 & 0.467 & 0.858 \\
\hline
0.7 & 0.20 & 0.100 & 1.0$\times10^{-3}$ & 0.0056 & $-$0.262 & 0.458 & 0.794 \\
0.7 & 0.10 & 0.098 & 4.8$\times10^{-4}$ & 0.0056 & $-$0.260 & 0.456 & 0.795 \\
\hline
\end{tabular}\end{center}
\end{table}

\begin{figure}[th]
\centering\leavevmode
\includegraphics[width=4.5in]{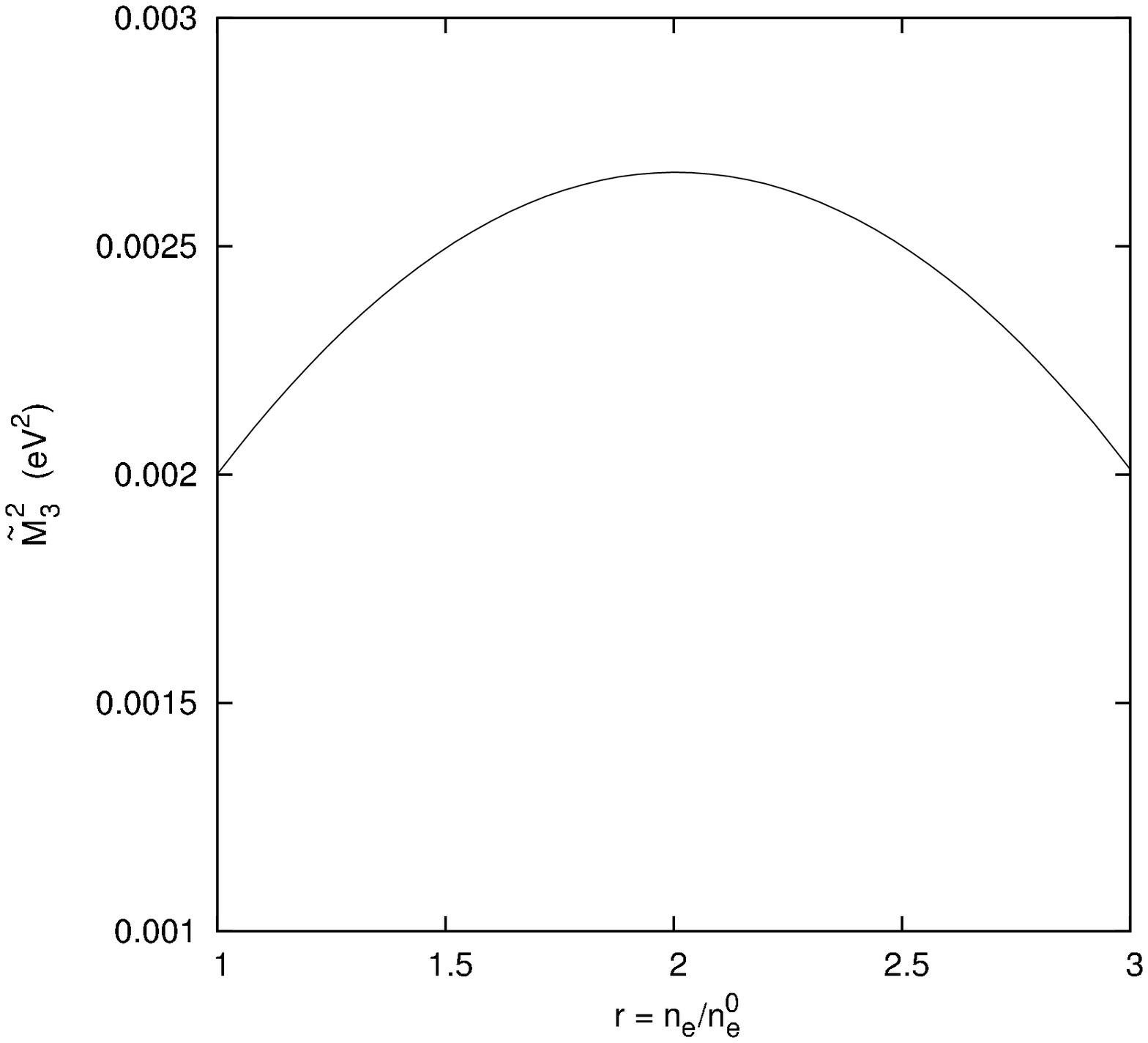}
\caption[]{Variation of $\tilde M_3^2$ (which determines the $\delta m^2$ for
atmospheric and long-baseline neutrinos) with electron number density
for the first example listed in Table~\ref{tab:one}. Similar behavior
is exhibited for the other examples.
\label{fig:fig3}}
\end{figure}

We note that there is some fine tuning required so that the K2K and
atmospheric neutrino mass-squared differences are small compared to
the LSND scale.  This amounts to requiring $|m_3 - M_{33}| M_{44}
\simeq M_{34}^2$; since $|m_3| \ll |M_{33}|$, this is approximately
equivalent to requiring $|M_{33}M_{44}| \simeq M_{34}^2$. Since the
MaVaN parameters scale similarly with density, this fine tuning is not
greatly upset by changes in density, {\it e.g.}, from the Earth's crust to
its core.  This is confirmed by Fig.~\ref{fig:fig3}, which shows the
value of $\tilde M_3^2$ for the first example in Table~\ref{tab:one}
as $r$ is increased from 1 to 3. Since $\tilde M_3^2$ and
$\tilde\theta$ do not vary greatly over the entire range of
Earth densities, the oscillation formulas in Sec.~2.2 should provide a
good approximation to the actual oscillation probabilities.
Furthermore, the model should provide good fits to the atmospheric
neutrino data over all zenith angles in which the neutrino path is
mostly in the Earth. The parameters $M_{33}, M_{34}$ and $M_{44}$ are
also changed very little by changes to the input values for $k$ (the
power that determines the density dependence of the MaVaN parameters);
$m_3$ increases by about 10\% if $k$ is decreased to $1/2$. Therefore
the general features of the model are insensitive to the exact density
dependence of the MaVaN effects, to changes in the matter density as
the neutrinos traverse the Earth, or to the precise value of the
core/crust density ratio.

The pathlength in Earth matter for atmospheric neutrinos is given by
\begin{equation}
L_m = \sqrt{R^2\cos^2\Theta + \sin^2\Theta(2R\epsilon-\epsilon^2)}
- (R-\epsilon)\cos\Theta \,,
\end{equation}
where $\Theta$ is the zenith angle (zero for downward events), $R$ is
the Earth's radius and $\epsilon$ is the detector depth (of order 1~km).
For comparison, the total path length is
\begin{equation}
L = \sqrt{R^2\cos^2\Theta + \sin^2\Theta(2R\epsilon-\epsilon^2)
+ 2R\delta + \delta^2} - (R-\epsilon)\cos\Theta \,,
\end{equation}
where $\delta$ is the height above the Earth's surface at which the
neutrino is created (of order 20~km).  If $L_m$ is a few oscillation
lengths or more, the oscillations average and the distinction between
$L_m$ and $L$ is inconsequential. However, when $L_m$ is significantly
different from $L$ and the oscillations do not average, the matter
mixing angles no longer accurately describe the oscillations. For
example, for horizontal events ($\Theta = \pi/2$) the total distance
is less than an oscillation length for the atmospheric $\delta m^2$
and $L_m/L \simeq \sqrt{\epsilon/(\delta+\epsilon)} \simeq \sqrt{1/20}
\simeq 0.22$.  For events with $\Theta$ near $\pi/2$, the difference
between $L_m$ and $L$ changes rapidly with $\Theta$; a detailed
analysis of atmospheric data is required to determine the precise
constraints on the MaVaN parameters.

If $\theta_a \ne \pi/4$, {\it i.e.}, the atmospheric neutrino mixing is not
maximal, then the maximum LSND amplitude for a given value of
$\theta_x$ can be larger, as given by the formula
\begin{equation}
\sin^22\theta_L^{max} = {(\sin^22\theta_L^{max})_{\theta_a = \pi/4}
\over 1-\cos2\theta_a}.
\label{eq:not-max}
\end{equation}
For example, for $\sin^22\theta_a = 0.9$ the maximum LSND amplitude
increases by about 50\%. This has the effect of shifting the solid
curves to the right in Figs.~\ref{fig:fig1} and \ref{fig:fig2} by the
same factor. Current data still excludes the model, and the allowed
parameters shift somewhat in the scenario where MiniBooNE sees a null
result, but the general features remain the same.

\subsection{Solar neutrinos}

The matter density in the sun is much higher than in the Earth, but
since in our model MaVaN effects only occur for $\nu_3$ and $\nu_4$,
$\delta M^2_{21}$ has approximately the same value as in the standard
MSW scenario for all densities. The effects of sterile mixing are also
small since $\nu_s$ couples only indirectly to the two lightest
states. We have checked numerically that the MSW probability in
Eq.~(\ref{eq:solar}) including all MaVaN and sterile mixing effects
gives the usual result to within about a percent or less for the
same values of $\theta_s$, $\theta_a$ and $\theta_x$. Also, we found
that the fraction of solar neutrinos that oscillate to steriles
(Eq.~\ref{eq:solar-sterile}) is less than 1\%, which is easily within
the range allowed by current data~\cite{bbmpy}.  Therefore the MaVaN
mass terms and sterile neutrino mixing have a negligible effect on
solar neutrino oscillations in our model.

A non-zero $\theta_x$ improves the fit to the intermediate-energy
solar neutrinos (compared to the two-neutrino case), at the expense of
a slightly worse fit to the low and high-energy solar
neutrinos~\cite{bmwLMA}. As noted previously, a value for
$\sin^22\theta_x$ as large as $\simeq 0.30$ is consistent with
combined fits to solar and KamLAND data~\cite{fogli} and does
not appear to be ruled out. The range $0.10 \lsim
\sin^22\theta_x \lsim 0.30$ that would be consistent with LSND and a
null MiniBooNE result is also consistent with the solar data.

\subsection{Vacuum constraints}

The constraints which apply to the experiments primarily in vacuum
(Bugey and CHOOZ) must also be checked for consistency with the
model. Since the vacuum $m_3^2$ is of order $\delta m^2_s$ and the
vacuum $m_4^2=0$, all of the oscillations in vacuum occur with $\delta
m^2$ values of order $\delta m^2_s$, not accessible in short
baseline experiments. Therefore both the Bugey and CHOOZ bounds are
avoided in this model.

We note that there are solutions to Eqs.~(\ref{eq:m33})-(\ref{eq:m34}) and
(\ref{eq:tilde}) with $m_4 = 0$ other than those listed in
Table~\ref{tab:one}. They have the same sign for $m_3$ and $M_{33}$,
with $m_3 \sim M_{ij}$, {\it i.e.}, all parameters involving $\nu_3$
and $\nu_4$ are of ${\cal O}(1)$~eV. In that case, there are vacuum
oscillations at short baseline due to $m_3^2 \sim 1$~eV$^2$. However,
such solutions would have $\bar\nu_e \to \bar\nu_e$ oscillations at
short baseline with approximate amplitude $\sin^22\theta_x$ which
are ruled out by the Bugey reactor experiment.

\section{Discussion}

We conclude by discussing some of the main features of our MaVaN model
with three active neutrinos and one sterile neutrino designed to explain
the LSND data and a null MiniBooNE result:

\begin{itemize}

\item The sterile mixing angle in the Earth's crust ($\theta$) is
approximately the same as the sterile mixing angle in the Earth's core
($\tilde\theta$); these angles must obey $\sin^2\theta \simeq
\sin^2\tilde\theta \lsim 0.10$ to agree with atmospheric neutrino
oscillation data.

\item A large part of the region in $\delta m^2_L$-$\sin^22\theta_L$
space consistent with the LSND/null-MiniBooNE region, can be
reproduced in this MaVaN model with $0.10 \lsim \sin^22\theta_x \lsim
0.30$ and $0.04 \lsim \sin^2\theta \lsim 0.10$. The significant size
of $\theta_x$ means that it should be detectable in proposed reactor
experiments with expected sensitivity $\sin^22\theta_x \ge 0.01$ where
most of the neutrino path is in Earth matter, such as Angra,
Braidwood, Daya Bay~\cite{white-paper}, or
KASKA~\cite{kaska}. However, Double-CHOOZ~\cite{doublechooz}, which
should be sensitive to $\sin^22\theta_x \ge 0.03$, would see a null
result since most of the neutrino path is in air (where the $\delta
m^2$ values are all of order $\delta m^2_s$).  The planned
long-baseline experiments MINOS~\cite{minos} and ICARUS~\cite{icarus},
sensitive to $\sin^22\theta_x \ge 0.05$ at the 90\%~C.L.\cite{lbl},
should also see a positive signal in the $\nu_\mu \to \nu_e$
appearance channel. The large value for $\theta_x$ might also be
detectable in future measurements of solar neutrinos~\cite{bmwLMA}.

\item The generic features of the model are relatively insensitive to
the precise density dependence of the MaVaN parameters. There is a
certain amount of fine tuning between vacuum and MaVaN parameters
required to achieve the relation $\delta m^2_a \ll \delta m^2_L$ in
the Earth, but the mass-squared differences for atmospheric neutrinos
are fairly stable under variations in the Earth density.

\item The {\it vacuum} neutrino masses of the active states 
are given by $m_1 = 0$, $m_2 =
\sqrt{\delta m^2_s} \simeq 0.009$~eV and $m_3 \simeq 0.006$~eV. If we
allow $m_1 = M_1$ to be non-zero, it can be of order $m_2$ and $m_3$
if there is a small upward shift in the eigenvalues $M_3^2$, $M_4^2$
and $\tilde M_3^2$ by the non-zero value of $m_1^2$. This shift does
not make an appreciable difference to the parameters in
Table~\ref{tab:one}. In this case there would be neither a hierarchy
{\it nor} a zero value required in either the vacuum masses of the
active neutrinos or in the
MaVaN couplings.

\item The values of $\delta m^2_a$ measured in the K2K and atmospheric
neutrino experiments do not have to be the same, since they are
separate inputs in determining the model parameters. The central value
in K2K is about 40\% higher than for atmospheric neutrinos (with
90\%~C.L.  uncertainties of order 50\%). Although currently not
significant, if the discrepancy between $\delta m^2$ values for K2K
and atmospheric neutrinos persists, it could easily be accommodated in
this model.

\item It has been shown that MaVaN terms that involve only $\nu_1$ and
$\nu_2$ can improve the fit to solar neutrino
data~\cite{solar-mavans}.  Diagonal and off-diagonal MaVaN couplings
must be introduced for $\nu_1$ and $\nu_2$; if there are no
MaVaN terms coupling $\nu_1$ or $\nu_2$ with $\nu_3$ or $\nu_4$,
then the phenomenology at the $\delta m^2_a$ and $\delta m^2_L$ scales
discussed in this paper is unaffected. The MaVaN couplings involving
$\nu_1$ and $\nu_2$ would need to be about three orders of magnitude
smaller than those for $\nu_3$ and $\nu_4$.

\item We have chosen to consider only MaVaN couplings that involve the
third and fourth generations, so that they do not impact the solar
neutrino $\delta m^2$ scale for Earth matter densities. Introducing
MaVaN terms that couple $\nu_1$ or $\nu_2$ to $\nu_3$ or $\nu_4$ can
also induce $\nu_\mu \to \nu_e$ oscillations at the LSND scale, but
then at solar densities this mixing upsets the value of the solar
$\delta m^2$. Although we have not performed an exhaustive parameter
search, MaVaN terms that lead to the appropriate LSND amplitude and
that also couple appreciably to $\nu_1$ and $\nu_2$ at Earth matter
densities appear to be problematic.

\item The MINOS experiment has recently found evidence for 
$\nu_\mu \to \nu_\mu$ oscillations at a baseline of 735 km~\cite{minos1}
with a $\delta m^2$ that is consistent with K2K and atmospheric data. 
Since the neutrino path in the MINOS experiment traverses matter of similar
density as that in the K2K experiment, our model also explains MINOS data.

\end{itemize}

In summary, we have presented a MaVaN model that can explain all
neutrino oscillation results, including LSND, and a null result for
$\nu_\mu \to \nu_e$ oscillations in MiniBooNE. There is no hierarchy
required in the vacuum masses of the active neutrinos, 
which are of ${\cal O}(10^{-2})$~eV, and
the density-dependent MaVaN parameters are all of ${\cal O}(1)$~eV for
the matter density of the Earth's crust. Active-sterile mixing is
small and is generated solely by MaVaN effects. Due to the large value
required for $\theta_x$, the model predicts visible oscillation
effects in underground reactor neutrino experiments such as Daya Bay
and Braidwood, but a null result in the mostly above ground
Double-CHOOZ experiment. Long-baseline experiments such as MINOS and
ICARUS should see sizeable $\nu_\mu \to \nu_e$ oscillations.

\section*{Acknowledgments}

We thank Janet Conrad, Bert Crawley, John Learned, Tom Meyer, Sandip
Pakvasa, Wesley Smith and Graham Wilson for conversations and
especially Jack Steinberger for correspondence and Renata Zukanovich-Funchal
for noting that we had omitted 
the neutral-current forward 
scattering contribution in Eq.~(\ref{eq:prop}). 
  VB thanks the Aspen
Center for Physics for hospitality during the completion of this
work. This research was supported by the U.S. Department of Energy
under Grants No.~DE-FG02-95ER40896 and DE-FG02-01ER41155, by the NSF
under CAREER Grant No.~PHY-0544278 and Grant No.~EPS-0236913, 
by the State of Kansas through the
Kansas Technology Enterprise Corporation and by the Wisconsin Alumni
Research Foundation.


\end{document}